\documentclass{aa}

\usepackage{graphics}

\def\gtaprx {\lower .1ex\hbox{\rlap{\raise .6ex\hbox{\hskip .3ex
             {\ifmmode{\scriptscriptstyle >}\else
                {$\scriptscriptstyle >$}\fi}}}
                \kern -.4ex{\ifmmode{\scriptscriptstyle \sim}\else
                {$\scriptscriptstyle\sim$}\fi}}}

\begin{document}

\thesaurus{}
\title{Millimeter observations of radio-loud active galaxies}

\author{Ilse M. van Bemmel\inst{1,2} \and Frank Bertoldi\inst{3}}

\institute{European Southern Observatory, Karl-Schwarzschildstr. 2,
           D--85748 Garching bei M\"unchen
        \and
        Kapteyn Astronomical Institute, P.O. Box 800, NL--9700~AV
           Groningen
        \and
        Max-Planck-Institut f\"ur Radioastronomie, Auf dem H\"ugel 69,
           D-53121 Bonn}

\offprints{Ilse van Bemmel (bemmel@astro.rug.nl)}

\date{Received date; accepted date}

\authorrunning{van Bemmel \& Bertoldi}
\titlerunning{Millimeter observations of radio-loud AGN}

\maketitle

\begin{abstract}

In order to study the nature of the far-infrared emission observed in
radio-loud active galaxies, we have obtained 1.2\,mm observations with
the IRAM 30\,m telescope for a sample of eight radio-loud active
galaxies. In all objects we find that the 1.2\,mm emission is
dominated by non-thermal emission. An extrapolation of the non-thermal
radio spectrum indicates that the contribution of synchrotron emission
to the far-infrared is less than 10\% in quasars, and negligible in
the radio galaxies. The quasars in the sample show signs of
relativistic beaming at millimeter wavelengths, and the quasar 3C\,334
shows evidence for strong variability. 


\end{abstract}

\section{Introduction}
Photometry from radio to infrared wavelengths has shown that
double-lobed radio-loud quasars show deep minima in their spectral energy
distributions at millimeter wavelengths (Antonucci et al. 1990), which
indicates that the radio emission arises from components which cannot
be related to the infrared emission. However, the nature of the
far-infrared emission observed in double-lobed radio sources remains
unclear. Most evidence (Haas et al. 1998, Polletta et al. 2000, and
references therein) points at a thermal nature of the infrared
emission, although a possible contribution from relativistically
beamed, non-thermal emission has not yet been carefully determined. 
Relativistic beaming can play an important role in the infrared
emission of quasars, in which the relativistic jet is oriented closer
to the line of sight (Barthel 1989). In radio-galaxies however,
beaming is thought to have a negligible effect (Hoekstra et al. 1997).  

The significance of beaming may be estimated by extrapolating the radio 
core spectra to infrared wavelengths. The non-thermal emission from
the radio lobes can be safely neglected, as these have no bulk
relativistic motions and thus show no 
beaming. However, the exact shape of the core spectrum is unknown at
high frequencies. 

In order to improve the quality of the extrapolation
of the non-thermal spectrum, we measured 1.2\,mm continuum fluxes for
eight 3CR objects, of which seven have been observed with ISOPHOT on
board ISO (Lemke et al. 1996; Kessler et al. 1996) and with the NRAO
Very Large Array (VLA) (van Bemmel et  al. 2000). We here compare the
1.2\,mm fluxes with the integrated radio and infrared data, and for the
quasars in the sample, with their radio core spectra. 

The general properties of our objects are listed in Table~1. The
sample consists of three radio-loud quasars (QSR), two broad-line
radio galaxies (BLRG), one narrow-line radio galaxy (NLRG) and the
radio structure 3C\,59. The latter was previously misidentified with
the Seyfert\,1 galaxy RBS\,0281. The maps of Meurs \& Unger (1991)
show three components: a north-western source (the original 3C\,59), a
central source associated with RBS\,0281, and a south-eastern
source. 3C\,59 has no optical counterpart, so it is either a lobe of
RBS\,0281, or a background core-dominated quasar. With ISOPHOT only
RBS\,0281 was observed. We shall refer to the Seyfert galaxy as
RBS\,0281, and to the north-eastern hotspot/background object as 3C\,59.  

%
%
\begin{table}[!ht]
\label{chartab}
\begin{center}
\leavevmode
\footnotesize
\begin{tabular}[h]{lccccc}
\hline \\[-8pt]
Name                & IAU      & $z$   & type & $L_{178}$ & size \\
                    & B1950    &       &      & [W/Hz]    & [\arcsec] \\
\hline \\[-8pt]
\object{3C\,33.1}   & 0106+729 & 0.181 & BLRG & 26.89  & 227   \\
\object{3C\,59}     & 0204+293 &   --  &  ?   & 24.01  & $\sim 4$   \\
\object{RBS\,0281}  & 0204+293 & 0.110 & Sey1 &   --   & $\sim 1$  \\
\object{3C\,67}     & 0221+276 & 0.310 & BLRG & 27.28  & 2.5  \\
\object{3C\,277.1}  & 1250+568 & 0.321 & QSR  & 27.24  & 1.7  \\
\object{3C\,323.1}  & 1545+210 & 0.264 & QSR  & 27.13  & 69   \\
\object{3C\,334}    & 1618+177 & 0.555 & QSR  & 27.88  & 58   \\        
\object{3C\,460}    & 2318+235 & 0.268 & NLRG & 27.08  & 8    \\
\hline \\[-8pt]
\end{tabular}
\end{center}
\caption{Basic properties of the observed objects. Types are: NLRG =
narrow-line radio galaxy, BLRG = broad line radio galaxy, QSR = radio
quasar, Sey1 = Seyfert 1, ? = unknown. We adopt the most common
classification in the literature.
$L_{178}$ assumes H$_0$\,=\,75\,km\,s$^{-1}$\,Mpc$^{-1}$,
q$_0$\,=\,0.5 and $F_{\nu}$\,$\propto$\,$\nu^{-1}$.
The last column lists the size of the radio structure associated with
the object, only the central component size is given for RBS\,0281.}
\end{table}
\normalsize

\section{Observations and data reduction}
\subsection{IRAM 30\,m}
Continuum observations at 1.2\,mm (250\,GHz) were obtained between 12 
and 15 December 1999 with the Max-Planck Millimeter Bolometer (MAMBO;
Kreysa et al. 1998) at the IRAM 30\,m telescope on Pico Veleta, Spain
(Baars et al. 1987). MAMBO is a 37-element bolometer array, sensitive
between 190 and 315\,GHz. The half peak sensitivity range is 210 --
290\,GHz, with an effective bandpass center, somewhat dependent on the
spectral slope of the observed emission, at 250\,GHz.  The beam for
the feed horn of each bolometer is matched to the telescope beam of
10.6\arcsec, and the bolometers are arranged in a hexagonal pattern
with a beam separation of 22\arcsec.  Observations were made in
standard on-off mode, with 2\,Hz chopping of the secondary reflector
by 32\arcsec. The pointing accuracy is typically 2\arcsec. The target
was centered on the central bolometer of the array, and after each 10
seconds of integration, the telescope was nodded so that the previous
``off'' beam becomes the ``on'' beam. Each scan of twelve 10 second
subscans lasts 3 minutes, of which 1 minute integration falls on the
sources, 1 minute off source, and 1 minute is used to move the
telescope and start the integration. Typically 15 to 20 of such scans
were performed for each object.  

Gain calibration was performed using observations of Mars, Uranus, and
Ceres, resulting in a flux calibration factor of 12500 counts per
Jansky, which we estimate to be accurate to 15\%. A sky opacity
correction factor was measured every 2 hours through total power sky
dips. 

The data were analyzed using the MOPSI software package (Zylka 1998).
For each bolometer the temporally correlated variation of the sky
signal (sky-noise) was computed using the signals of neighbouring
bolometers. The correlated noise is iteratively determined for each
channel and subtracted. 

Due to unintentional mispointing, for 3C\,59 and RBS\,0281 the targeted
positions differ from the radio positions in the maps of Meurs \&
Unger (1991, see also Table~2). A signal is measured at the targeted
position toward RBS\,0281, which however is 10\arcsec\ off the radio peak, so
that the millimeter flux from the peak radio position could well be
higher than observed. Toward 3C\,59, a 7\,mJy signal was picked up by
an off-center channel, closely corresponding to the position of
component~D in the Meurs \& Under radio map. Since the chances of
observing a background source are very small, we assume that the
off-center channel detects 3C\,59, but also here the peak
flux could be higher than the observed flux. Since
the MAMBO fluxes toward RBS\,0281 and 3C\,59 are uncertain, we will
not make use of them in the analysis. They will be re-measured in a
future observing campaign. 

\begin{table}[!t]
\label{chartab}
\footnotesize
\begin{center}
\leavevmode
\begin{tabular}[h]{llll}
\hline \\[-8pt]
Object 	& Position of	& RA (2000)	& DEC (2000)	\\
\hline \\[-8pt]
3C\,59	& MAMBO target	& 02 07 09.6	& 29 31 24	\\
	& radio peak	& 02 07 10.1	& 29 31 45.1	\\
	& detection 	& 02 07 09.7	& 29 31 44	\\
\hline \\[-8pt]
RBS\,0281 & MAMBO target& 02 07 02.3	& 29 30 55.1	\\
	& radio peak	& 02 07 02.2	& 29 30 46.8	\\
\hline \\[-8pt]
\end{tabular}
\end{center}
\caption{Observed positions for 3C\,59 and RBS\,0281: central channel
	position and the peak flux positions in the Meurs \& Unger
	(1991) maps are given. For 3C\,59 we also list the position at
	which the signal is detected.}
\end{table}
\normalsize

\subsection{Radio and infrared data}
We collected radio fluxes for our objects from the NASA/IPAC
Extragalactic Database (NED). Where available, we adopt those given at
178\,MHz, 1.4\,GHz, 2.7\,GHz, and 4.9\,GHz.  For the three quasars
additional fluxes from the unresolved core are available at 4.9, 15,
25 and 43\,GHz from a previous programme (van Bemmel et al. 2000,
hereafter BBG). For RBS\,0281 only one radio point was found.

The infrared data were obtained with ISOPHOT, using the P1, P2, C100 
and C200 detectors in raster-mapping mode. A detailed description of 
the data reduction is given in BBG, including a list of the resulting
flux densities. ISOPHOT data are available at 60, 90 and 160\,$\mu$m
for all objects except 3C\,59. For 3C\,33.1 and RBS\,0281 we have
additional ISOPHOT data at 12 and 25\,$\mu$m. For 3C\,323.1 an upper
limit at 10\,$\mu$m was obtained by Rieke \& Low (1972).

\section{Results}
%
%
\begin{table*}[!ht]
\label{chartab}
\footnotesize
\begin{center}
\leavevmode
\begin{tabular}[h]{lcccccccc}
\hline \\[-8pt]
Name      &178~MHz&1.4~GHz&2.7~GHz&4.9~GHz&250~GHz \\
          & (Jy) & (Jy) & (Jy) & (mJy) & (mJy) \\       
\hline \\[-8pt]
3C\,33.1  & 11.4$\pm$0.9 & --            & --            & 820$\pm$62  & 6$\pm$2 \\
3C\,59    & 5.3$\pm$1.0  & 1.60$\pm$0.03 & --            & 670$\pm$87  & (7$\pm1$) \\
RBS\,0281 & --           & 0.026$\pm$0.002 & --            & --          & ($>3$) \\
3C\,67    & 9.0$\pm$1.1  & 3.10$\pm$0.09 & --            & 860$\pm$113 & 10.3$\pm$0.7 \\
3C\,277.1 & 8.9$\pm$0.9  & 2.50$\pm$0.09 & 1.56$\pm$0.06 & 880$\pm$90  & 20.6$\pm$1.5 \\
3C\,323.1 & 10.0$\pm$1.3 & 2.45$\pm$0.12 & 1.50$\pm$0.08 & 840$\pm$112 & 13.5$\pm$0.9 \\
3C\,334   & 10.9$\pm$0.8 & 2.20$\pm$0.11 & 1.00$\pm$0.05 & 500$\pm$76  & $<$4 \\
3C\,460   & 8.1$\pm$1.0  & 1.60$\pm$0.08 & 0.90$\pm$0.05 & 440$\pm$59  & 4.3$\pm$1.2 \\
\hline \\[-8pt]
\end{tabular}
\end{center}
\caption{NED archive radio fluxes $S_{\nu}$ and newly obtained MAMBO
250\,GHz fluxes for our sample. Additional data can be found in van
Bemmel et al. (2000). Brackets indicate an off-center detection.} 
\end{table*}
\normalsize

Table~3 lists the radio and millimeter flux densities. The radio
fluxes are integrated fluxes, which include the lobes and other
extended structures. The millimeter fluxes only include the part of
the object within the IRAM~30\,m telescope beam. The resulting
spectral energy distributions (SED) are plotted in Fig.~1, along with
the radio flux densities of the unresolved core emission for the
quasars. 

The 1.2\,mm flux densities are always lower than the infrared and the
integrated radio flux densities, consistent with observations by
Antonucci et al. (1990). They found that quasars show a minimum in
their SEDs at millimeter wavelengths.  We obtained similar results for
radio galaxies. The millimeter flux densities are even lower than
the fluxes expected from an extrapolation of the integrated radio
spectra, which we show as dashed lines in Fig.~1. For this
extrapolation we fitted a power law to the radio data with
$\nu>1$\,GHz. In cases where a flux was available at only one
frequency, we use the average slope of the other objects, excluding
3C\,59 and RBS\,0281 though.

Half of the objects in our sample have radio sizes larger than the
10.6\arcsec\ beam of the IRAM 30\,m telescope at 1.2\,mm. In these
cases a direct comparison with the radio data may not be meaningful,
unless we assume that the radio lobes are not contributing to the
millimeter emission (see section 4.3). For the quasars, core radio
fluxes are available at even higher resolution (1--2\arcsec). The core
radio SED shows no obvious relation to the millimeter fluxes, except
maybe for 3C\,323.1, where the core and millimeter fluxes
show a typical self-absorbed synchrotron spectrum. 

To test whether the millimeter flux could arise from a cold, thermal
component that also gives rise to the 160\,$\mu$m emission, we fit an
optically thin grey-body spectrum to the far-infrared flux, using a
dust emissivity index $\beta=1.6$. This value is an average of
observed values in nearby active and normal galaxies.
To be conservative,
we adopted dust temperatures of 20\,K, which is even colder than the
coldest dust found in active galaxies with ISO (e.g. Siebenmorgen et
al. 1999). The  temperature is typical for dust in Galactic clouds and
cirrus. The dust masses implied by such a cold component range from a
few $10^7$ to several $10^8$\,M$_{\odot}$. From the extrapolated
grey-body spectra it appears that the millimeter flux is dominated by
non-thermal emission. In 3C\,460, e.g., the 1.2\,mm flux could be
dominated by thermal dust emission if $\beta<1.4$ or $T<16$, but then
the implied dust mass would be larger than $5\times10^8$\,M$_{\odot}$.
Such a large dust mass is hard to reconcile with 3C\,460 being an
elliptical galaxy. There is only circumstantial evidence for a very
cold dust component ($T<15$\,K) in galaxies. Starbursts and active
galaxies do not show very cold dust emission at millimeter wavelengths.

3C\,334 is probably a variable millimeter source. Previous
observations with the Owens Valley Radio Observatory found a 3\,mm
flux density of $37\pm 2$\,mJy (van Bemmel et al. 1998). Assuming
$F_{\nu}$\,$\propto$\,$\nu^{-1}$, a 1.2\,mm flux of 15\,mJy would be
expected, much higher than our upper limit of 3\,mJy. 3C\,334 is known
to be variable at 4.85\,GHz, with variations of order 10\% over a few
decades (van Bemmel et al. 1998). Variability tends to increase at
higher frequencies, but one would not expect a dramatic change in flux
such as implied by the OVRO and MAMBO measurements. 

%
%
\begin{figure*}
  \resizebox{18cm}{!}{\includegraphics{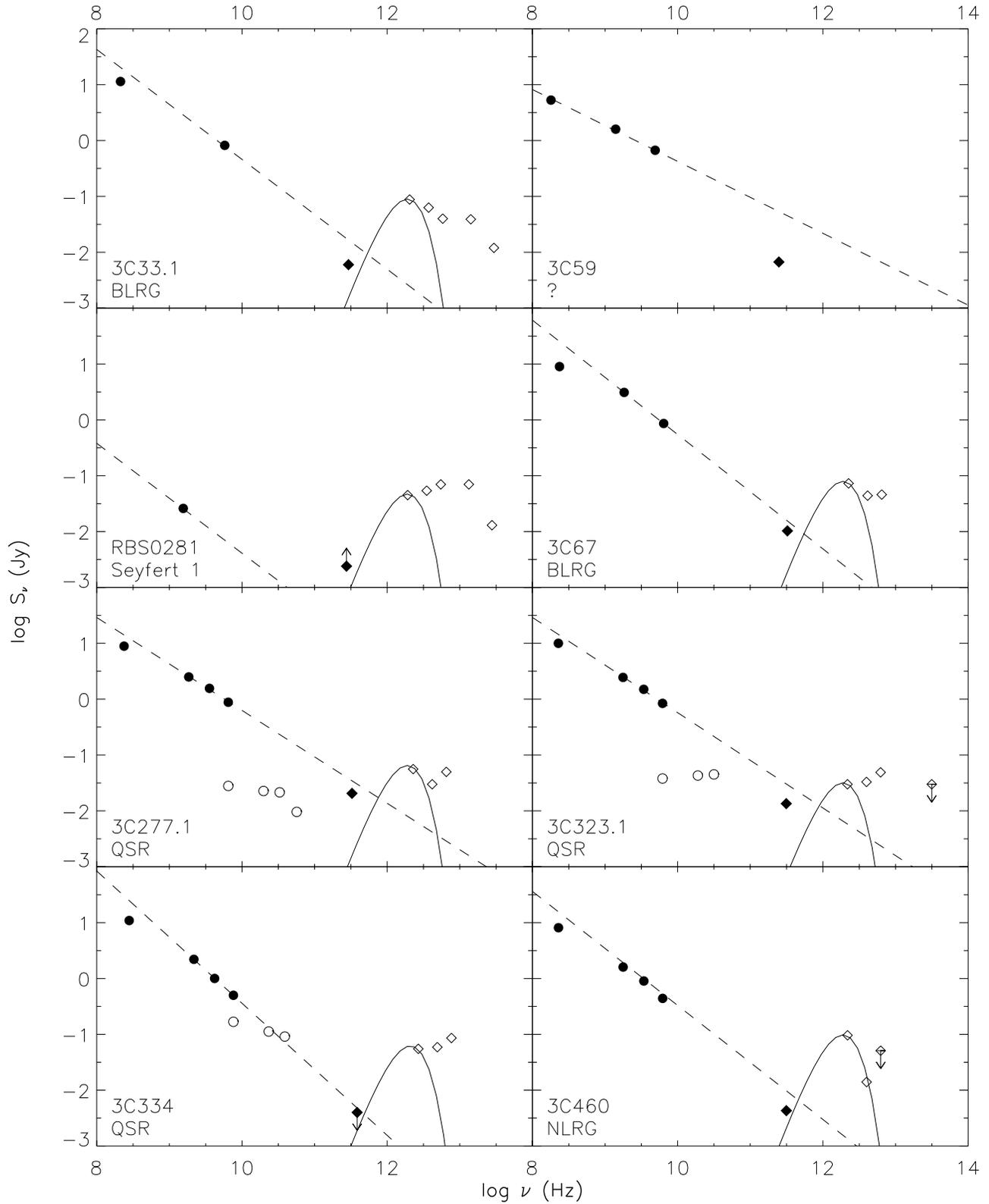}}
  \hfill
  \parbox[b]{18cm}{
  \caption{Rest frame spectral energy distributions of the objects
  observed with MAMBO. For 3C\,59 we assume $z$\,=\,0. 
  Filled circles show the total integrated radio fluxes,
  open circles show the unresolved core radio fluxes, the filled
  diamond the MAMBO 1.2\,mm fluxes, and open diamonds are ISOPHOT
  fluxes. The solid lines represent grey-body $T$\,=\,20\,K spectra,
  matching the 160\,$\mu$m fluxes. The dashed lines are power law fits
  to $>$\,1\,GHz integrated radio data.}
}\label{sed}
\end{figure*}
\normalsize
\subsection{Luminosities and spectral indices}
%
%
\begin{table}[!t]
\label{chartab}
\footnotesize
\begin{center}
\leavevmode
\begin{tabular}[h]{lcccc}
\hline \\[-8pt]
Name    & $\log L_{250}$ & $\alpha^{178}_{1.4}$ & $\alpha^{1.4}_{4.9} $
& $\alpha^{4.9}_{250}$ \\
\hline \\[-8pt]
3C\,33.1 & 23.61   & --    & --    & -1.28 \\
3C\,59   &  --	   & -0.61 & -0.64 & (-1.18)\\
RBS\,0281& ($>$22.86)   & --    & -- & ($>$-0.42)$^*$\\
3C\,67   & 24.33   & -0.52 & -1.02 & -1.15 \\
3C\,277.1& 24.67   & -0.62 & -0.83 & -0.97 \\
3C\,323.1& 24.30   & -0.68 & -0.85 & -1.07 \\
3C\,334  & $<$24.46  & -0.78 & -1.18 & $<$-1.25\\
3C\,460  & 23.82   & -0.79 & -1.03 & -1.2 \\
\hline \\[-8pt]
QSR average& 24.49 & -0.69 & -0.96 & -1.02 \\
RG average & 23.92 & -0.65 & -1.03 & -1.21 \\
\hline \\[-8pt]
\end{tabular}
\end{center}
\caption{Monochromatic 250\,GHz luminosities (in W\,Hz$^{-1}$) and
radio spectral indices. Brackets indicate an off-center detection.
($^*$) for RBS\,0281 $\alpha^{1.4}_{250}$ is given
instead of $\alpha^{4.9}_{250} $.}
\end{table}
\normalsize

Adopting $H_0$\,=\,75, $q_0$\,=\,0.5, and $k\Lambda$\,=\,0, we computed 
luminosity densities in order to compare the millimeter luminosities
of quasars and radio galaxies (Table~4). 3C\,59 is excluded, because
its redshift has not yet been determined. Although we find that the 
infrared luminosities are comparable between the classes, there is
evidence that the quasars are brighter at 1.2\,mm than the radio
galaxies. At 178\,MHz the quasars are only marginally brighter. 

The sample is too small to draw firm conclusions, but it appears that
the stronger millimeter emission from quasars may be due to a beamed
component. When comparing the integrated flux at 178\,MHz and 1.4\,GHz,
the quasars are a factor 1.5 brighter than the radio galaxies. In case
of isotropic synchrotron emission over the whole radio spectrum, this
factor should be constant. However, it appears to increase toward
higher frequencies; at 250\,GHz the quasars are about three times
brighter than the radio galaxies. This indicates that we observe an
additional component, which is unlikely to be due to dust emission,
since that would be optically thin and therefore also visible in the
radio galaxies. For synchrotron emission, the excess can only be
beamed emission, which causes a natural anisotropy. It would be
desirable to confirm this trend with a larger sample. 

There is clear evidence of spectral steepening of the radio spectrum
towards millimeter wavelengths. The average spectral index for the entire
sample ranges from $-0.7$ at the lowest frequency to $-1.2$ at millimeter 
wavelengths (Table~4). Spectral steepening occurs in most objects in
our sample,
irrespective of their size or the resolution of the observations.

\section{Discussion}
\subsection{Nature of the millimeter and far-infrared emission}

We find that the 1.2\,mm flux densities are well below those expected
from an extrapolation of the total radio flux densities. If cold dust
gives rise to the millimeter emission, then the millimeter flux
densities are expected to lie above the extrapolated radio spectra,
and the grey-body extrapolation should fit the observed flux. In our
data the opposite is true, and therefore the millimeter emission is
probably dominated by synchrotron emission, except maybe for 3C\,460,
where this case is not so clear.  

Since the far-infrared fluxes are  more than an order of magnitude
higher than the millimeter fluxes, the far-infrared emission is likely
to arise from dust. The clear minimum of the SED at millimeter
wavelengths excludes that the far-infrared emission is the
continuation of the radio synchrotron spectrum of the radio lobes or
the core. 

If the far-infrared emission were due to synchrotron radiation, it
would have to arise from a very young electron population. The younger
the electrons, the closer they must be to the core, so that their
emission would be even more strongly beamed than synchrotron emission
at millimeter wavelengths. As a consequence, the quasars should be
brighter than radio galaxies at infrared wavelengths, for which there
is no evidence in our data. Instead we find that in the infrared the
quasars and radio galaxies are comparably bright, while at 1.2\,mm the
quasars are clearly brighter than the radio galaxies.  

The possible contribution of relativistically beamed synchrotron
emission to the far-infrared emission is estimated by extrapolating
the radio SED. This yields on average a less than 10\% contribution
in quasars, and a less than 1\% contribution in radio galaxies. 
Although in 3C\,323.1 the core radio spectrum would extrapolate well 
to the observed far-infrared fluxes, the 1.2\,mm emission falls below
this extrapolation, indicating a turnover in the core spectrum, which
excludes a significant contribution of non-thermal emission to the
far-infrared. 

An earlier study of the infrared emission from radio-loud active
galaxies indicated that quasars are significantly brighter at
60\,$\mu$m than radio galaxies (Heckman et al. 1992). This does not
contradict what we find from our sample. The objects in
the Heckman sample have an average redshift $\sim$\,$0.5-0.9$, thus in
their rest frame the emission emerges at $\sim$\,$30-40\,\mu$m. If the
infrared emission arises from a circumnuclear torus, the dust is likely
to be optically thick up to $60$\,$\mu$m (Pier \& Krolik 1992, Granato
\& Danese 1994), and the emerging flux depends on the orientation of
the torus. According to these models, the observed flux could vary by
orders of magnitude, depending on the torus' optical depth. However,
ISOPHOT studies (BBG) show no conclusive evidence that quasars are
intrinsically brighter in the far-infrared than radio galaxies, when
they are matched in radio power and redshift.  

\subsection{Relativistic beaming in quasars}

Unified models for radio-loud AGN (Urry \& Padovani 1995, Barthel
1989) suggest that quasars are oriented with their jets closer to the
line of sight than radio galaxies. This implies that any beamed
component is more evident in quasars, but also that isotropic
emission, such as optically thin dust emission, should not differ
among the types. Our observations confirm this picture, showing that
quasars are more luminous at millimeter wavelengths, where we expect
relativistic beaming, whereas they do not differ in their far-infrared
luminosity, which is due to isotropic cold dust emission, and therefore
unaffected by beaming. The luminosity ratio QSR/RG increases with
frequency, which can only naturally be explained by beaming. 

If this trend is confirmed in larger samples, millimeter observations
in combination with radio observations could provide a direct measure
of the orientation of a source. E.g. for a quasar and a radio galaxy
of comparable 178\,MHz power, the radio galaxy provides the unbeamed
radio SED that can be subtracted from the quasar SED. The remaining
emission is then due to beaming, and the strength of this beamed
component depends directly on the viewing angle of the source.

\subsection{Hotspot and lobe emission}
For three of our objects, the radio structures are larger than the
beam size of the IRAM~30\,m telescope, so that the hotspots fall
outside the central bolometer channel. We made no attempt to observe
the hotspots separately. 

The radio fluxes for all objects are integrated fluxes, including the
lobe emission. The MAMBO fluxes are also integrated fluxes for the
small objects, but core fluxes for the objects with larger radio sizes
than the IRAM beam. If the millimeter emission would be dominated by
the lobe emission in all objects, there should be a clear turnover in
the SEDs of the large objects, where the lobes are not observed with
MAMBO. On the other hand, if the core dominates the millimeter
emission in all objects, the SEDs 
should be comparable, irrespective of object radio size. We observe no
clear difference between the radio--millimeter SED of large and small
objects, which seems to indicate that the core is dominating the millimeter
emission. This does not imply that the lobes do not
emit any millimeter emission, e.g. in Cygnus~A the hotspots have been
clearly detected with SCUBA at 850\,$\mu$m (Robson et al. 1998).  

However, there might be a relation between the luminosity density of
the lobes at 1.2\,mm and the size of the radio structure. Small radio
sources are known to have much flatter spectra (Murgia et al. 1999)
and thus can have much stronger lobes. In our MAMBO observations, all
small sources are unresolved and thus we cannot tell which component
is dominating. If the lobes dominate in small objects and the core
dominates in larger ones, the expected difference in spectral indices will
not be visible. We always observe the dominating regions, i.e. the lobes in
the small radio sources and only the core in the larger ones. Judging
from the radio core fluxes in 3C\,277.1, the core is not the dominant
1.2\,mm source here. The same can be true for 3C\,67 and 3C\,460,
which are also small radio sources. The lobe emission will dilute the
amount of beaming observed in small objects, thus for a proper estimate
of the amount of beaming the objects should have comparable radio
sizes and be larger than the IRAM\,30m telescope beam.
\vspace{-2mm}
\subsection{Variability}
We find evidence for variability in 3C\,334. Since we have only one
observation for each object, variability cannot be ruled out for the
other objects. Previous SCUBA observations of the other quasars (BBG)
are consistent with the MAMBO detections, confirming the spectral
steepening and the thermal nature of the infrared emission.
For the radio galaxies there is remarkably little dispersion in
the observed spectral slopes. Thus, variability must be small in all
other objects.

\vspace{-1mm}
\section{Conclusions}
The main conclusions, drawn from our continuum observations at 1.2\,mm
of eight radio-loud active galaxies, can be summarized as follows.

\begin{itemize}
\vspace{-3mm}
\item The millimeter emission is dominated by
non-thermal processes, whereas the far-infrared emission must
be thermal.

\item Any thermal contribution to the millimeter emission arising from
cold dust is estimated to be less than 15\%, except for 3C\,460, where
the case is not clear.

\item An extrapolation of the non-thermal radio spectrum
to far-infrared wavelengths shows that the contribution 
of non-thermal emission at 160\,$\mu$m is less than 10\%.

\item There is evidence that at 1.2\,mm the quasar emission is
stronger, and therefore possibly more beamed, than that from radio
galaxies. 

\item The far-infrared luminosity on the other hand, does not differ
between quasars and radio galaxies. Most likely because it arises from
optically thin thermal emission, but optical thickness effects can
play a role up to 100\,$\mu$m in the rest frame of the objects. 

\item Radio and millimeter observations provide an interesting test
for unification models of radio-loud active galaxies, in that they can
provide a measure of the amount of beamed emission. Millimeter
observations, in combination with radio observations, of a larger
sample of radio-loud active galaxies are needed to verify this.

\end{itemize}

\begin{acknowledgements}
Thanks to Bob Fosbury and Peter Barthel for their motivation and help
on the manuscript.
Thanks to Alessandra Bertarini for assisting with the observations,
and to the referee, Neal Jackson, for comments
which greatly improved the manuscript. Special thanks to the MPIfR
bolometer team for providing MAMBO and support, and to R. Zylka for
writing the MOPSI data reduction package.

NED is operated by the Jet Propulsion Laboratory, California Institute
of Technology, under contract with the National Aeronautics and Space
Administration. 
ISO is an ESA project with instruments funded by ESA Member States
(especially the PI countries: France, Germany, the Netherlands and the
United Kingdom) and with the participation of ISAS and NASA. 
\end{acknowledgements}

\end{document}